\documentclass[a4paper]{jpconf}
\usepackage{graphicx}
\usepackage{amsmath}
\usepackage{amssymb}
\usepackage{fancyvrb}
\pdfoutput=1
\newcommand{\be}{\begin{equation}}
\newcommand{\ee}{\end{equation}}

\begin{document}
\title{WimPyDD: an object-oriented Python code for WIMP-nucleus scattering direct detection in virtually any scenario}

\author{Gaurav Tomar$^a$, Injun Jeong$^b$, Sunghyun Kang$^b$, Stefano Scopel$^b$}

\address{$^a$Physik-Department, Technische Universit\"at M\"unchen, \protect\\~James-Franck-Stra\ss{}e, 85748 Garching, Germany}
\address{$^b$Department of Physics, Sogang University, Seoul, Korea, 121-742}
\ead{physics.tomar@tum.de}

\begin{abstract}
We introduce WimPyDD, a modular, object–oriented and customisable Python code that accurately predicts the expected WIMP-nucleus scattering rates in WIMP direct–detection experiments including the response of the detector. WimPyDD utilises the framework of Galilean–invariant non–relativistic effective theory, allowing to handle an arbitrary number of effective operators, and can perform the calculation of the excepted rate in virtually any scenario, including inelastic scattering, WIMPs with an arbitrary spin, and a generic velocity distribution in the Galactic halo. The power and flexibility of WimPyDD is discussed in some explicit examples.
\end{abstract}

\section{Introduction}
Weakly Interacting Massive Particles (WIMPs) provide the most popular candidate of dark matter (DM) and there are various direct detection (DD) experiments searching for WIMP-nuclei scattering events in underground laboratories. Basically, the calculation of DD signals depends on different inputs coming from particle physics, nuclear physics, and astrophysics. Additionally, the realistic prediction from experiments require energy resolution, efficiency and the specific response to the target materials such as quenching.   
We introduce WimPyDD~\cite{Jeong:2021bpl}, a customisable, object-oriented Python code that allows to take all such inputs into account and predicts accurately the expected rate in an experiments virtually in any scenario, including WIMPs of arbitrary spin, inelastic scattering, and a generic velocity distribution of WIMP in the Galactic halo.

WimPyDD works in the framework of non-relativistic effective field theory (NREFT) for spin-1/2~\cite{Anand:2013yka} as well as arbitrary spin of DM~\cite{Gondolo:2020wge}. While calculating the expected rate in an experiment WimPyDD factorizes the calculation into three parts namely i) the Wilson coefficients that depends on the model parameters; ii) a response function that depends on experimental inputs and nuclear physics; iii) the halo function that contains inputs from astrophysics and depends on WIMP velocity distribution. It is only at the last step that all of these factorised calculations are combined to calculate the expected number of events.
\section{Scattering rate}
The expected number of WIMP events in a DD experiment in
the interval of visible energy $E_1^{\prime}\le E^{\prime}\le
E_2^{\prime}$ can be written in the form~\cite{DelNobile:2013cva},

\begin{equation}
  R_{[E_1^{\prime},E_2^{\prime}]}(t)=\frac{\rho_\chi}{m_\chi}\int_{v_{T^*}}^{\infty}
  \sum_T dv {\cal R}_{T,[E_1^{\prime},E_2^{\prime}]}(v)\eta(v,t),
  \label{eq:r_eta}
  \end{equation}

\noindent where $\rho_\chi$ is the local WIMP density, $m_\chi$ the WIMP mass, $T$ is the sum over the nuclear targets, and, 

\begin{equation}
v_{T^*}\equiv\left\{\begin{aligned}
\sqrt{\frac{2\delta}{\mu_{\chi T}}} ,\quad
\text{if}\;\delta> 0\\
0,\quad
\text{if}\; \delta\le 0\, ,
\end{aligned}\right.
\label{eq:vstar}
\end{equation}  
where $\delta$ is WIMP mass splitting and $\mu_{\chi T}$ is WIMP-nucleus reduced mass. The halo function in the expression above is,
\begin{equation}
 \eta(v,t)
 =\int_{v}^{\infty}\frac{f(v^{\prime},t)}{v^{\prime}}\,dv^{\prime},
\label{eq:eta}  
\end{equation}
which depends on the WIMP speed distribution through $f(v,t)\equiv 1/(4\pi) \int d\Omega
{v^\prime}^2f(\vec{v}^\prime,t)$. In WimPyDD the speed distribution is utilised in terms of a superposition of $N_s$ streams,
\begin{equation}
f(v,t)=\sum_k^{N_s} \lambda_k(t) \delta(v-v_k).
\end{equation}
Using the expression above the expected rate in an experiment can be written as,
\begin{equation}
  R_{[E_1^{\prime},E_2^{\prime}]}(t)=\frac{\rho_\chi}{m_\chi}\sum_{k=1}^{N_s}
  \delta\eta_k(t) \int_{v_{T^*}}^{v_i}
  \sum_T dv {\cal R}_{T,[E_1^{\prime},E_2^{\prime}]}(v),
  \label{eq:r_eta_streams}
  \end{equation}
where $\delta\eta_k(t)=\lambda_k(t)/v_k$. Since most of the DD experiments put upper bound on the time average of $R_{[E_1^{\prime},E_2^{\prime}]}(t)$, we have,

\begin{equation}
S^{(0)}_{[E_1^{\prime},E_2^{\prime}]}\equiv \frac{1}{T_0}\int_0^{T_0}
R_{[E_1^{\prime},E_2^{\prime}]}(t)dt,
\label{eq:s0}
\end{equation} 
with $T_0=1$ year, utilising which we arrived on the master formula used by WimPyDD,
\begin{eqnarray}
  &&S^{(0)}_{[E_1^{\prime},E_2^{\prime}]}=\frac{\rho_{\chi}}{m_{\chi}}\sum_{k=1}^{N_s} \delta\eta_k^{(0)}\times \sum_T\sum_{ij}\sum_{\tau\tau^{\prime}} c_j^{\tau}(w_i,q_0)c_{k}^{\tau^\prime}(w_i,q_0^{\prime})  \nonumber\\
  &&  \left \{ \left [\bar{{\cal R}}^0_{T,[E_1^{\prime},E_2^{\prime}]}\right ]_{jk}^{\tau\tau^{\prime}}(E_R)
  +(\frac{v_k^2}{c^2}-\frac{\delta}{\mu_{\chi T}})\left [
  \bar{{\cal R}}^1_{T,[E_1^{\prime},E_2^{\prime}]}\right ]_{jk}^{\tau\tau^{\prime}}(E_R) \right .\nonumber\\
&& \left .  -\frac{m_T}{2\mu_{\chi T}^2}
  \left [\bar{{\cal R}}^{1E}_{T,[E_1^{\prime},E_2^{\prime}]}\right ]_{jk}^{\tau\tau^{\prime}}(E_R)-\frac{\delta^2}{2 m_T}
  \left[\bar{{\cal R}}^{1E^{-1}}_{T,[E_1^{\prime},E_2^{\prime}]}\right ]_{jk}^{\tau\tau^{\prime}}(E_R)\right \}^{E_R^{max}(v_k)}_{E_R^{min}(v_k)},\nonumber \\
 && {} \label{eq:rate_rbar_inelastic}
\end{eqnarray}
\noindent where $\{f(E_R)\}_{E_1}^{E_2} \equiv f(E_2)-f(E_1)$ and integrated response functions $\bar{{\cal R}}_T$ are explicitly provided in~\cite{Jeong:2021bpl}.
\section{WimPyDD}
WimPyDD is available at the following webpage,\\
\begin{Verbatim}[frame=single,xleftmargin=1cm,xrightmargin=1cm,commandchars=\\\{\}]
  https://wimpydd.hepforge.org
\end{Verbatim}

\noindent where the code and additional explanations are available. The latest version of WimPyDD can be installed using,
\begin{Verbatim}[frame=single,xleftmargin=1cm,xrightmargin=1cm,commandchars=\\\{\}]
git clone https://phab.hepforge.org/source/WimPyDD.git
\end{Verbatim}
or can be directly downloaded from WimPyDD's homepage. This will create a folder WimPyDD. There are standard libraries required by WimPyDD such as matplotlib, numpy, scipy, and pickle. To use WimPyDD the user has to import it using, for instance,
\begin{Verbatim}[frame=single,xleftmargin=1cm,xrightmargin=1cm,commandchars=\\\{\}]
import WimPyDD as WD
\end{Verbatim}
\subsection{Signal routines}
There are three main routines in WimPyDD that calculate the differential cross section $(d\sigma/dE_R)_T$, the differential rate $dR/dE_R$, and the integrated rate of Eq.~(\ref{eq:r_eta}).

To calculate the differential cross section $(d\sigma/dE_R)_T$ (in
cm$^2$/keV) use,

\begin{Verbatim}[frame=single,xleftmargin=1cm,xrightmargin=1cm,commandchars=\\\{\}]
  WD.dsigma_der(element_obj,hamiltonian_obj,mchi,v,er,**args)
\end{Verbatim}
\noindent with \verb |mchi| the WIMP mass $m_\chi$ in GeV,
\verb |v| the WIMP incoming speed $v$ in km/sec,
\verb |er| the recoil energy in keV, and where \verb |element_obj| defines a nuclear target $T$ while \verb |hamiltonian_obj| defines an effective Hamiltonian. To evaluate the differential rate $dR/dE_R$ use,

\begin{Verbatim}[frame=single,xleftmargin=1cm,xrightmargin=1cm,commandchars=\\\{\}]
  WD.diff_rate(target_obj,hamiltonian_obj,mchi,\\
  energy,vmin,delta_eta, **args)
\end{Verbatim}
where \verb |target_obj| is either a single element or a combination of elements with stoichiometric coefficients, \verb |vmin| and \verb |delta_eta| are two arrays containing the $v_k$ and $\delta\eta_k^{(0)}$ in km/s and (km/s)$^{-1}$, respectively. The routine 
\verb |streamed_halo_function| is provided in WimPyDD for calculating them. The expected rate including the response of the detector is calculated by the routine,
\begin{Verbatim}[frame=single,xleftmargin=1cm,xrightmargin=1cm,commandchars=\\\{\}]
  WD.wimp_dd_rate(exp_obj, hamiltonian_obg, vmin,\\
  delta_eta, mchi,**args)
\end{Verbatim}

\noindent where \verb |exp_obj| is instantiated by the \verb |experiment| class
and contains all the required information about the experimental
set-up, including the energy bins where the rate is integrated. There are two arrays produced as an output of this routine with the centers
of energy bins and the corresponding expected rates calculated using
Eq.~(\ref{eq:rate_rbar_inelastic}).
\begin{figure}
\begin{center}
  \includegraphics[width=0.6\columnwidth]{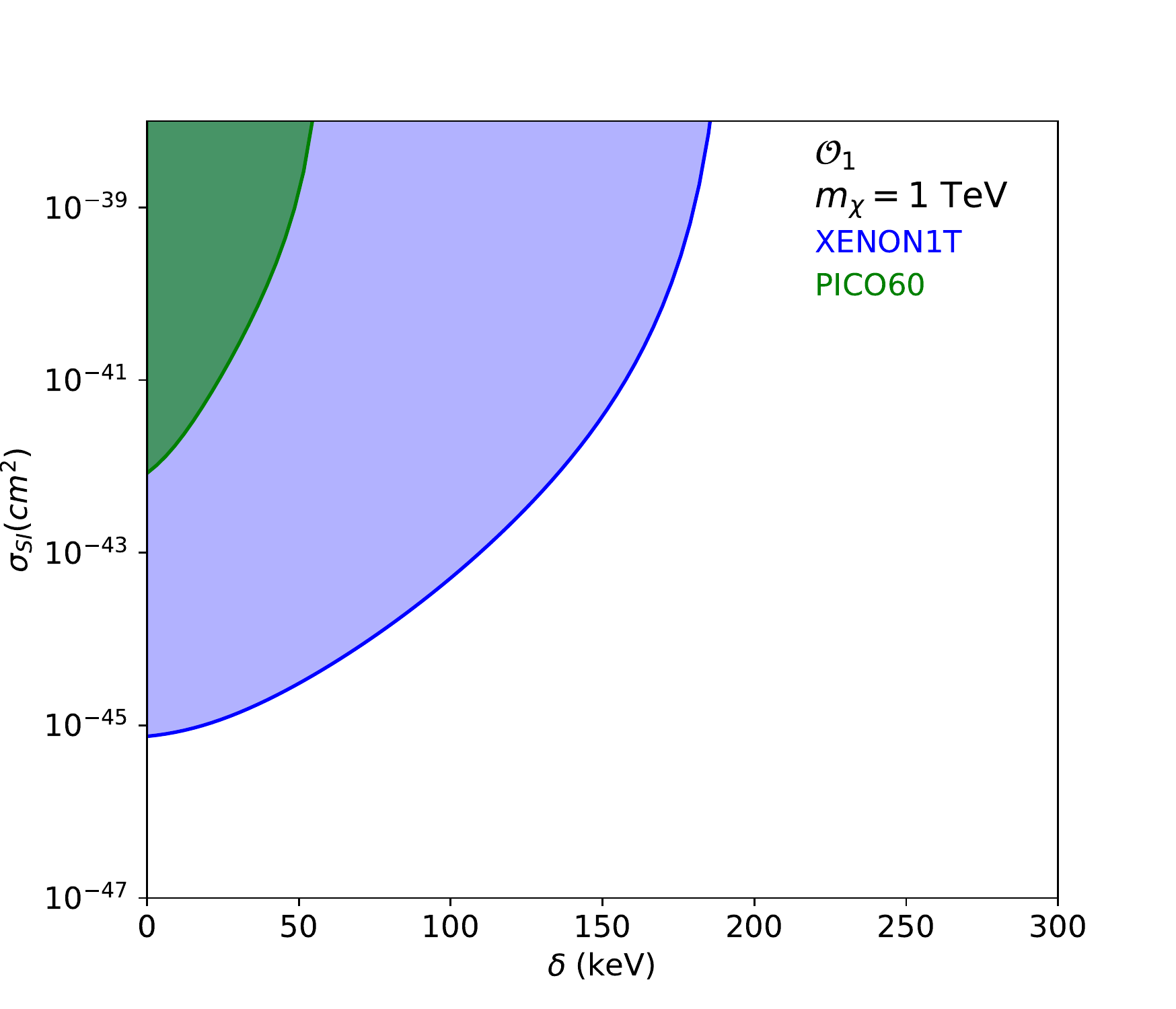}
  \end{center}
\caption{Exclusion limits on WIMP-nucleon SI scattering cross-section as a function of mass splitting $\delta$ at 90\% confidence for XENON1T~\cite{XENON:2018voc} and PICO60~\cite{PICO:2019vsc} experiments. We considered $m_\chi=1$ TeV and a WIMP escape velocity $v_{\rm esc}=550$ km/s in the Galactic rest frame.}
\label{fig:inelastic}
\end{figure}
\subsection{Example}
In the present section we introduce the potentialities of WimPyDD by explicitly showing example of WIMP-nucleus inelastic scattering.
\subsubsection{Inelastic scattering}
\hfill\\
Let’s set up the following effective Hamiltonian for spin-independent (SI) interaction $\mathcal{O}_1$,
\begin{equation}
{\cal H}=c_1^0{\cal O}_1 t^0+c_1^1{\cal O}_1
t^1= \sqrt{\frac{\sigma_{SI}\pi}{\mu^2_{\chi p}\hbar^2 c^2}}[(1+r){\cal O}_1 t^0+(1-r){\cal O}_1 t^1].
\label{eq:hamiltonian_example1}
\end{equation}
We utilised $c^0_1=c^p_1(1+r)$ and $c^1_1=c^p_1(1-r)$ with $r=c^n_1/c^p_1$, where $c^p_1$ and $c^n_1$ are the WIMP couplings to the proton and the neutron respectively. We replaced $c^p_1$ following,
\begin{equation}
	c^p_1=\sqrt{\frac{\sigma_{SI}\pi}{\mu^2_{\chi p}\hbar^2 c^2}}.
\end{equation}	
In the expression above $\sigma_{SI}=(c_1^p \mu_{\chi p})^2/\pi$ is the WIMP-nucleon cross-section while $\mu_{\chi p}$ is the WIMP-nucleon reduced mass. This is implemented in WimPyDD as following, 
\begin{Verbatim}[frame=single,xleftmargin=1cm,xrightmargin=1cm,commandchars=\\\{\}]
def wc(mchi,cross_section,r):\\
    hbarc2=0.389e-27\\
    mn=0.931\\
    mu=mchi*mn/(mchi+mn)\\
    cp=(np.pi*cross_section/(mu**2*hbarc2))**(1./2.)\\
    return cp*np.array([1.+r,1.-r])\\
    \\	
SI=WD.eft_hamiltonian('SI',\{1: wc\})
\end{Verbatim}
As discussed the WimPyDD code provides a simple routine \verb |mchi_vs_exclusion| that allows to estimate the upper bound on any parameter for which exclusion cross-section is chosen. The cross-section can be factorised from the rate by simply comparing the output of 
\verb |wimp_dd_rate| with $\sigma_{SI}=1$ cm$^2$ to the experimental upper bounds on the count rate in each energy bin of the considered experiment and taking the most stringent constraint. 
In \verb |mchi_vs_exclusion| we did not pass the $\sigma_{SI}$ argument so that it is set to 1 by default. The mass splitting $\delta$ goes as an input to the routine \verb |mchi_vs_exclusion| and the exclusion cross-section is scanned in terms of $\delta$ for XENON1T~\cite{XENON:2018voc} and PICO60~\cite{PICO:2019vsc} experiments as following,
\begin{Verbatim}[frame=single,xleftmargin=1cm,xrightmargin=1cm,commandchars=\\\{\}]
delta_vec=np.linspace(0.1,300,100)\\
WD.load_response_functions(WD.XENON_1T_2018,SI)\\
WD.load_response_functions(WD.PICO60_2019,SI)\\
vmin,delta_eta=\\
WD.streamed_halo_function(n_vmin_bin =1000, v_esc_gal=550)\\

sigma_xenon=[WD.mchi_vs_exclusion(WD.XENON_1T_2018, SI, vmin,\\
delta_eta, delta=delta, mchi_vec=np.array([1000]), \\
j_chi=0.5)[1] for delta in delta_vec]\\
sigma_pico=[WD.mchi_vs_exclusion(WD.PICO60_2019, SI, vmin,\\
delta_eta, delta=delta, mchi_vec=np.array([1000]), \\
j_chi=0.5)[1] for delta in delta_vec]\\

pl.plot(delta_vec[:64],sigma_xenon[:64],'-b')
pl.plot(delta_vec[:20],sigma_pico[:20],'-g')
\end{Verbatim}
In Fig.~\ref{fig:inelastic} the exclusion plot for the cross-section $\sigma_{SI}$ is
plotted as a function of the mass splitting $\delta$ for a WIMP escape
velocity $v_{esc}$ = 550 km/s in the Galactic rest frame. We see that the
current XENON1T and PICO60 constraints require $\delta$ to be $\sim$185 keV
and $\sim$55 keV, respectively.


\section*{References}
\begin{thebibliography}{9}

\bibitem{Jeong:2021bpl}
I.~Jeong, S.~Kang, S.~Scopel and G.~Tomar,
[arXiv:2106.06207 [hep-ph]].

\bibitem{Anand:2013yka}
N.~Anand, A.~L.~Fitzpatrick and W.~C.~Haxton,
Phys. Rev. C \textbf{89}, no.6, 065501 (2014)
doi:10.1103/PhysRevC.89.065501
[arXiv:1308.6288 [hep-ph]].
  
\bibitem{Gondolo:2020wge}
P.~Gondolo, S.~Kang, S.~Scopel and G.~Tomar,
Phys. Rev. D \textbf{104}, no.6, 063017 (2021)
doi:10.1103/PhysRevD.104.063017
[arXiv:2008.05120 [hep-ph]].

\bibitem{DelNobile:2013cva}
E.~Del Nobile, G.~Gelmini, P.~Gondolo and J.~H.~Huh,
JCAP \textbf{10}, 048 (2013)
doi:10.1088/1475-7516/2013/10/048
[arXiv:1306.5273 [hep-ph]].
  
\bibitem{XENON:2018voc}
E.~Aprile \textit{et al.} [XENON],
Phys. Rev. Lett. \textbf{121}, no.11, 111302 (2018)
doi:10.1103/PhysRevLett.121.111302
[arXiv:1805.12562 [astro-ph.CO]].

\bibitem{PICO:2019vsc}
C.~Amole \textit{et al.} [PICO],
Phys. Rev. D \textbf{100}, no.2, 022001 (2019)
doi:10.1103/PhysRevD.100.022001
[arXiv:1902.04031 [astro-ph.CO]].
\end{thebibliography}
\end{document}